\documentclass[conference]{IEEEtran}
\IEEEoverridecommandlockouts

\usepackage{graphicx}
\usepackage{flafter}
\usepackage{cite}
\usepackage{graphicx}
\usepackage{amsmath,amssymb,amsfonts}
\usepackage{algorithmic}
\usepackage{graphicx}
\usepackage{textcomp}
\usepackage{xcolor}
\usepackage{tikz}
\usepackage[algo2e,ruled,linesnumbered,boxed,commentsnumbered]{algorithm2e}

\usepackage{latexsym}
\usepackage{adjustbox}
\usepackage{enumitem}
\usepackage{dcolumn}
\usepackage{booktabs}
\usepackage{amsmath}
\usepackage{multirow}
\usepackage{graphicx}
\usepackage{amssymb}
\usepackage{pifont}
\usepackage{amsmath,array,graphicx}
\usepackage{kantlipsum}
\def\BibTeX{{\rm B\kern-.05em{\sc i\kern-.025em b}\kern-.08em
    T\kern-.1667em\lower.7ex\hbox{E}\kern-.125emX}}
\usepackage{url}

\usepackage{breakurl}
\usepackage[breaklinks]{hyperref}
\usepackage{rotating}

\begin{document}


\title{Machine Learning Research Towards Combating \mbox{COVID-19}: Virus Detection, Spread Prevention, \mbox{and Medical Assistance}}

\author{
    \IEEEauthorblockN{Osama Shahid\IEEEauthorrefmark{1}, Mohammad Nasajpour\IEEEauthorrefmark{1}, Seyedamin Pouriyeh\footnote{Corresponding author}\IEEEauthorrefmark{1}, Reza M. Parizi\IEEEauthorrefmark{2}, Meng Han\IEEEauthorrefmark{1}, Maria Valero\IEEEauthorrefmark{1}, \\Fangyu Li \IEEEauthorrefmark{3},  Mohammed Aledhari\IEEEauthorrefmark{4},  Quan Z. Sheng\IEEEauthorrefmark{5}}
    \IEEEauthorblockA{\IEEEauthorrefmark{1} Department of Information Technology, Kennesaw State University, Marietta, GA, USA
    \\ \{oshahid1, mnasajp1\}@students.kennesaw.edu, \{spouriye, mhan9, mvalero2\}@kennesaw.edu}
    \IEEEauthorblockA{\IEEEauthorrefmark{2} Department of Software Engineering and Game Development, Kennesaw State University, Marietta, GA, USA   \\rparizi1@kennesaw.edu}
    
 \IEEEauthorblockA{\IEEEauthorrefmark{3}Department of Electrical and Computer Engineering, Kennesaw State University, Marietta, GA, USA
    \\ fli6@kennesaw.edu}

       \IEEEauthorblockA{\IEEEauthorrefmark{4}Department of Computer Science, Kennesaw State University, Marietta, GA, USA
    \\  maledhar@kennesaw.edu}
   
   \IEEEauthorblockA{\IEEEauthorrefmark{5}Department of Computing, Macquarie University, Sydney, Australia
    \\michael.sheng@mq.edu.au}
\thanks{ \scriptsize \noindent Corresponding author: Seyedamin Pouriyeh (email: spouriye@kennesaw.edu).}   
}

\maketitle
\begin{abstract}
\mbox{COVID-19} was first discovered in December 2019 and has continued to rapidly spread across countries worldwide infecting thousands and millions of people. The virus is deadly, and people who are suffering from prior illnesses or are older than the age of 60 are at a higher risk of mortality. Medicine and Healthcare industries have surged towards finding a cure, and different policies have been amended to mitigate the spread of the virus. While Machine Learning (ML) methods have been widely used in other domains, there is now a high demand for ML-aided diagnosis systems for screening, tracking, and predicting the spread of \mbox{COVID-19} and finding a cure against it. 
In this paper, we present a journey of what role ML has played so far in combating the virus, mainly looking at it from a screening, forecasting, and vaccine perspectives. We present a comprehensive survey of the ML algorithms and models that can be used on this expedition and aid with battling the virus. 

\end{abstract}

\section{Introduction}
In December 2019, a novel severe contagious respiratory syndrome coronavirus 2, which is a type of Severe Acute Respiratory Syndrome (SARS-CoV-2) virus called \mbox{COVID-19}, was discovered in Wuhan, China \cite{latif2020leveraging}. \mbox{COVID-19} virus is airborne and can easily spread and infect people \cite{morawska2020airborne}. According to the Centers for Disease Control and Prevention (CDC) \cite{CDCC19}, the infected people show a range of symptoms like dry cough, shortness of breath, fatigue, losing the sense of taste and smell, diarrhea, and congestion. Infected patients can also present fever episodes. 
Strangely enough, some patients who have contracted the virus might not even show any of the aforementioned symptoms \cite{nishiura2020estimation}. They can feel completely normal carrying the virus and continuing to spread the disease without knowing \cite{nishiura2020estimation}. As \mbox{COVID-19} has a rapid nature of spreading, the World Health Organization (WHO) declared it as a global pandemic in March 2020 \cite{al2020asymptomatic}. At the time of writing this paper (i.e., September 2020), the total number of confirmed \mbox{COVID-19} cases worldwide was over 32 million \cite{googlenews}. To tackle this outbreak, scientists in different research communities are seeking a wide variety of computer-aided systems such as the Internet of Things\cite{nasajpour2020internet},  Machine Learning (ML) or Deep Learning (DL) techniques\cite{vaishya2020artificial, alimadadi2020artificial}, Big Data\cite{bragazzi2020big}, and Blockchain\cite{chang2020can} that can assist with overcoming the challenges brought by \mbox{COVID-19}. These technologies can be used for controlling the spread of the virus,  detecting the virus, or even designing and manufacturing a vaccine or drug to combat it. 


There were two epidemics in the past from the coronavirus family including Severe Acute Respiratory Syndrome (SARS-CoV)\cite{darnell2004inactivation} and Middle Eastern Respiratory Syndrome (MERS)\cite{willman2019comparative}. SARS-CoV is a respiratory virus that was transmissible from person to person and it was first identified in 2003. The virus had over 8,000 confirmed cases worldwide during its course which affected over 26 countries \cite{sarsnews}. MERS is also a respiratory virus with similar symptoms of SARS-CoV.

ML, as a subset of Artificial Intelligence (AI), has shown a lot of potentials in many industries like retail \cite{jia2013retail}, banks \cite{chitra2013data}, healthcare \cite{chen2017disease,clemente2020smart}, pharmaceuticals \cite{ekins2016next}, and many more \cite{el2015machine}. ML techniques can be programmed to imitate human intelligence. For example, in the healthcare industry, ML techniques can be trained and used towards medical diagnosis \cite{alpaydin2020introduction}. 
ML models have been vastly trained over a dataset consisting of medical images like Computed Tomography (CT) Scan, Magnetic Resonance Imaging (MRI), or X-Ray to detect anomalies \cite{wong2019artificial01, ards2019artificial}. Its classification models can be expanded in diverse areas including cancer \cite{kourou2015machine}, diabetes \cite{zou2018predicting}, fatty liver \cite{wu2019prediction}, etc. As an example, breast cancer can be diagnosed with a prediction accuracy of 97.13\% \cite{asri2016using} using ML models.

During previous epidemics, ML techniques have been widely implemented in order to assist healthcare authorities for better actions regarding the diseases \cite{ahmad2018interpretable}. For example, Sandhu et al. \cite{sandhu2016intelligent} proposed an ML model that utilizes GPS technology along with cloud computing power and Google Maps to represent potentially infected patients and provide an alternative route for uninfected users resulting in potentially mitigating the spread. The model reaches the classification accuracy of 80\% in re-routing away from infected patients.
In another study, Choi et al. \cite{choi2017large} used ML models for sentimental analysis to review public overreaction appearing in media articles and social media platforms. This type of in-depth analysis can rapidly monitor the public reaction. It can also aid policymakers in taking the right actions in reducing fear and distress from the public regarding MERS.

ML has also been widely used in order to improve clinical decision-making regarding the current \mbox{COVID-19} pandemic \cite{debnath2020machine}. Researchers, using ML algorithms and clustering technique, are able to forecast the spread in provinces \cite{hu2020artificial,allam2020artificial}. ML methods of image classification are used by the scientific community to help in diagnosing the deadly virus \cite{ozkaya2020coronavirus}. With the objective of finding a cure for the virus, ML algorithms are used to evaluate how dependable are off-the-counter drugs may be used to help infected patients \cite{heiser2020identification}.

The aforementioned examples show the potential of ML in the detection, diagnosis and prediction of viruses. This paper largely provides a survey reviewing the research that has been dedicated by the scientific community in using ML technology in combating \mbox{COVID-19}. In particular, we investigate the role of ML in \textit{detecting or screening}, \textit{forecasting}, and \textit{medical assistance} for the virus.

The remainder of the paper is organized as follows. In Section \ref{sec:screening}, we review the role of ML in detecting and the screening process of \mbox{COVID-19}. We study the use of different ML techniques regarding four major sections for diagnosing and screening including Medical Imaging, Chatbot, and Artificial Intelligence of Things (AIoT). In Section \ref{sec:predicting}, the importance of disease contamination and its exposure to others are discussed regarding the use of ML for tracking and predicting the spread of \mbox{COVID-19}. This section is mainly divided into three parts reviewing preventing the spread, contact tracing, and forecasting. Similarly, Section \ref{sec:medicalAssistance} reviews the need for medical assistance during the pandemic and how ML technology can be integrated. Understanding the virus and developing a drug or vaccine using ML techniques are discussed in this section as well. Finally, we discuss, outline future work, and conclude in sections \ref{sec:FutureWorks} and \ref{sec:Conclusion} respectively.



\section{ML Techniques towards Detecting \& Screening \mbox{COVID-19}} \label{sec:screening}

Detecting either a symptomatic or asymptomatic disease in an early stage could be highly effective in order to start the process of treatment. 
Regarding the \mbox{COVID-19}, not only it helps to avoid the spread of contamination, but also it is cost-beneficial. 
The standard method of diagnosing \mbox{COVID-19} is to conduct Reverse-Transcription Polymerase Chain Reaction (RT-PCR) test \cite{corman2020detection}. The RT-PCR is a swab test that is used to detect nucleic acid from \mbox{COVID-19} in the upper and lower respiratory system. At the beginning of the pandemic, the sensitivity of the RT-PCR test could show negative for patients who were later in-fact confirmed positive, hence the reporting of false-negative rates was high \cite{xie2020chest}. There was also a concern about having a shortage and a limited number of tests, and the high-cost factor of producing and conducting them \cite{gollier2020group}. It is important to explore alternative methods of diagnosing \mbox{COVID-19} that would speed-up the process \cite{pham2020artificial}. To help overcome the challenge presented in diagnosing \mbox{COVID-19}. ML models and algorithms have shown promising results in different stages of \mbox{COVID-19} \cite{lalmuanawma2020applications}. 
In general, ML techniques have been widely utilized in the healthcare domain and similarly, it can be used towards analyzing data and diagnosing \mbox{COVID-19} using medical imaging which includes X-Ray and CT Scan images \cite{ozkaya2020coronavirus}. In this section, we review different ML techniques that have been used for screening and diagnosis of \mbox{COVID-19}. Moreover, we discuss other ML-based tools including Chatbots and Artificial Intelligence of Things (AIoT).


\subsection{Medical Imaging}
Diagnosing \mbox{COVID-19} is one of the most important parts of dealing with the disease. As a result of low access and high possibility of false-negative results to the RT-PCR kits, there is an essential need for using other approaches such as medical images analysis for accurate and reliable screening and diagnosis in \mbox{COVID-19}\cite{yang2020patients}. In general, analyzing medical imaging modalities such as chest X-ray and CT-Scan have key contributions in confirming the diagnosis of \mbox{COVID-19} as well as screening the progression of the disease \cite{kassani2020automatic}. Different ML techniques that incorporate X-ray and CT-Scan image processing approaches could help physicians and healthcare professionals as a better way for diagnosis and understanding of the progression of the \mbox{COVID-19} disease.


\subsubsection{X-ray}

\begin{figure*}[!tb]
\centering
    \includegraphics[scale=.34]{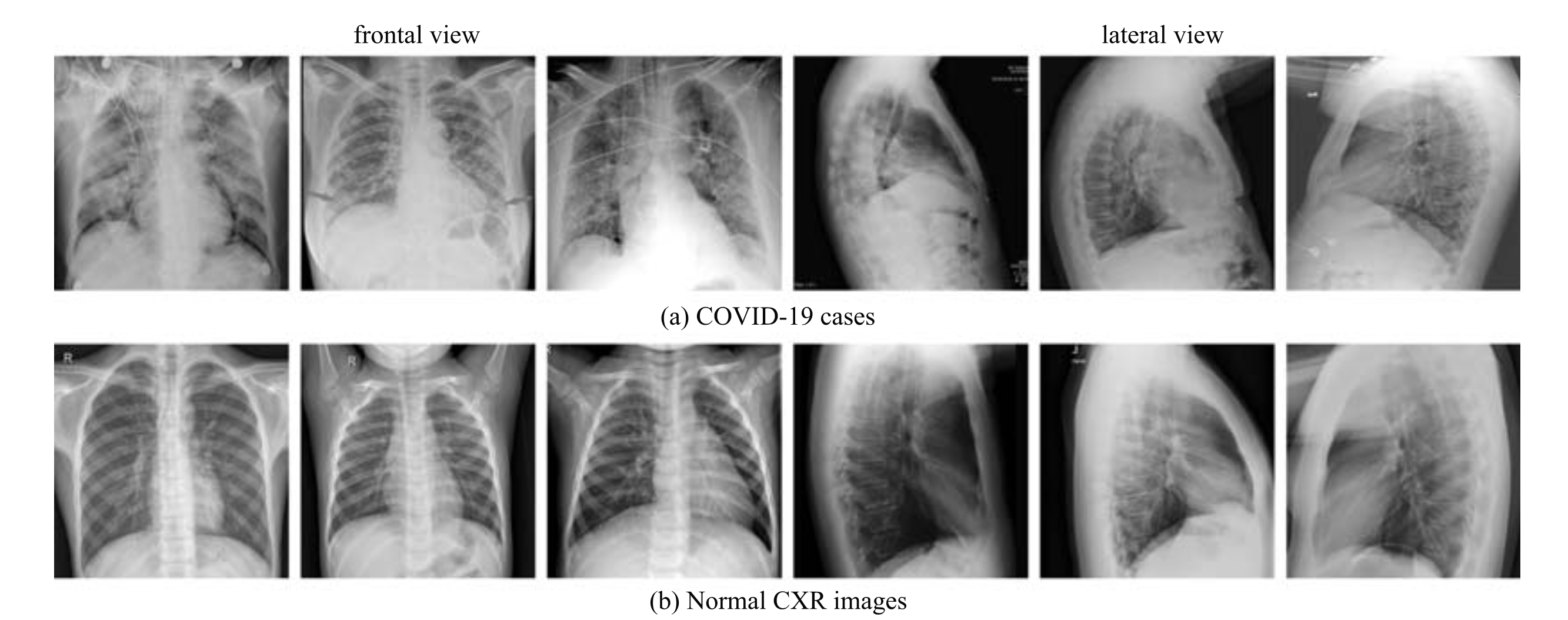}
    \caption{Chest X-Ray (CXR)  images of \mbox{COVID-19} infected people versus uninfected people \cite{chen2020survey}.}
    \label{image:CXRimage}
\end{figure*}

During this pandemic, chest imaging can be an important part of the \mbox{COVID-19} in early stage of detection. Classifying patients rapidly is what is expected from these approaches. Within the categorization of medical imaging, Chest X-Ray (CXR) was recommended to be implemented as the first medical imaging regarding \mbox{COVID-19} by the Italian Society of Radiology (SIRM) \cite{giovagnoni2020facing}. Figure \ref{image:CXRimage} demonstrates the CXR images from infected and normal people. According to Cozzi et al. \cite{cozzi2020chest}, CXR has a sensitivity of 67.1\% which can be first implemented in special cases including assisting radiologists with better \mbox{COVID-19} cases identification and fast treatment assigning to the patient. Additionally, CXR is inexpensive and secure because of minimizing the risk of contamination which makes a safer workplace for healthcare workers as well. 

In order to decrease the amount of work by radiologists, ML techniques can be assigned to classify patients with respect to \mbox{COVID-19}. To do that, researchers are mostly focused on the ML classification models such as Support Vector Machine (SVM), Convolutional Neural Networks (CNN), DL. One approach \cite{hassanien2020automatic} implemented X-Ray in order to classify the lung lesions (caused by \mbox{COVID-19}) with Multi-level Threshold (MT) process and SVM model. Within this model, firstly, the lung image contrast will be enhanced. Secondly, the image will be reduced into specific sections (using MT) to avoid duplication of work on uninfected areas. Lastly, the SVM model classifies the sections of the lung with respect to the predefined healthy lungs. Sethy et al. \cite{sethy2020detection} developed a platform using a variety of Deep Convolutional Neural Networks (DCNN) models classifying within the SVM with two different datasets in order to detect \mbox{COVID-19} cases based on the related CRX image. Similarly, \cite{castiglioni2020artificial} proposed a DCNN model using the data gathered from two hospitals in Italy to represents the importance of AI in the detection of \mbox{COVID-19}.

Zhang et al. \cite{zhang2020covid} trained ML models over a large viral pneumonia dataset of CXR images to detect anomalies. They tested their model on a completely different dataset that has \mbox{COVID-19} CXR images. This is done as one of the symptoms of \mbox{COVID-19} can be pneumonia \cite{CDCC19}. The results are impressive as the model performs well when tested on the \mbox{COVID-19} dataset with the Area Under the Curve (AUC) of 83.61\%. It is even more impressive as the model was trained on a different dataset and yet performed well. 
Similarly, Wang et al. \cite{wang2020covid} utilized  \textit{COVIDx} dataset, a publicly available dataset consists of \mbox{COVID-19}, pneumonia and non-\mbox{COVID-19} pneumonia-related X-ray images. The authors used this data to train their model for detection of \mbox{COVID-19}, the Deep Neural Network (DNN) is referred to as \textit{COVID-Net} showing promising results in diagnosing infected patients. 
Apostolopoulos et al. \cite{apostolopoulos2020covid} used transfer learning approaches like feature extraction and fine-tuning of CNN based models and trained and tested over similar datasets achieving a prediction accuracy up to almost 98\%. They demonstrated that implementing transfer learning can have a significant improvement in results. Most ML classifiers are trained and tested to achieve high prediction accuracy of \mbox{COVID-19}; however, it is also important to quantify the uncertainty that could exist by using such classifiers as a primary medium of diagnosis. An approach to validate the ML prediction of diagnosis in CXR images was reviewed by Ghoshal et al.\cite{ghoshal2020estimating}.
It exploited a Bayesian Deep Learning classifier to estimate the model uncertainty.
The result analysis displays a strong correlation between uncertainty and accuracy of prediction, which means that the higher the uncertainty outcome, the more reliable the prediction accuracy. 

Many other ML models are utilized for detecting \mbox{COVID-19}, and a subset of them are presented in Table~\ref{tab:xray}.
The references are both published papers and also papers that are yet to be peer-reviewed. All references that are presented in the table show high performance towards detecting predicting \mbox{COVID-19} through CXR images.


\renewcommand{\arraystretch}{1.2}
\begin{table*}[ht]
\centering
  \caption{\label{tab:xray} ML Research done towards diagnosing \mbox{COVID-19} using X-RAY (CXR) datasets.}
 \begin{adjustbox}{width=\textwidth}
 \begin{tabular}{llll}
 \hline
 \textbf{Reference} & \textbf{Dataset} & \textbf{Methods} & \textbf{Remarks}  \\ [0.5ex] 
 \hline

\cite{apostolopoulos2020covid} & Multiple Datasets that include 448 confirmed \mbox{COVID-19} images \textit{source: Github} & DL - CNN, Feature Extraction (various models) & Various model performance comparison  \\ 
\midrule

\cite{ghoshal2020estimating} & 68 \mbox{COVID-19} Chest X-ray images and 5873 Pneumonia images \textit{source: Github, Kaggle} & Bayesian Deep Learning Classifiers & Using Transfer learning approach with the classifier to estimate uncertainty \\  
\midrule

 \cite{farooq2020covid} & \multirow{12}{*}{COVIDx} & Fine Tuning - ResNet Model & Model achieves high accuracy for multi-class classification \\ \cline{3-4}\cline{1-1}
\cite{hemdan2020covidx} & & Multiple models VGG19, MobileNet & Test and train multiple image classifiers to obtain the highest accuracy identifying the virus; \\

 & & & VGG19 and DenseNet have a high accuracy score \\ 

\cline{3-4}\cline{1-1}
\cite{hirano2020vulnerability} &  & Multiple Image Classification Models & Demonstrate how DNNs are vulnerable to a universal adversarial perturbation causing failure \\

 &  &  & in classification tasks, fine-tuning may be able to improve this \\ 

\cline{3-4}\cline{1-1}
\cite{afshar2020covid} & & Capsule Network-based framework - COVID-CAPS & An alternative modeling framework that has decent performance with low trainable parameters  \\ \cline{3-4}\cline{1-1}
\cite{sarker2020covid} & & DenseNet-121 & The model is explained well, to test its robustness authors perform multi-class classification and perform k-fold validation \\ \cline{3-4}\cline{1-1}
\cite{ucar2020covidiagnosis} & & SqueezeNet-Bayesian based model & The authors classify input of X-Ray images of Normal, \mbox{COVID-19}, \\

& & &Pneumonia using data augmentation and fine-tuning techniques \\

\cline{3-4}\cline{1-1}
\cite{khobahi2020coronet} & & CoroNet -  & Semi-supervised learning technique based on AutoEncoders \\ \cline{3-4}\cline{1-1}
\cite{luz2020towards} & & EfficientNet & The proposed approach is able to produce a model with high quality with high accuracy \\ \cline{3-4}\cline{1-1}
\cite{wang2020covid} & & COVID-Net & COVID-Net is publicly accessible for scientists to further build and improve the network to achieve high accuracy  \\ \cline{3-4}\cline{1-1}
\cite{narin2020automatic} & CXR images of 50 \mbox{COVID-19} patients, and 50 Normal CXR images & Three different CNN models (InceptionV3, ResNet50, Inception-ResNetV2) & ResNet50 provides the highest classification. \\ 
\midrule
\cite{maghdid2020diagnosing} & 170 Chest X-Ray images of 45 patients from 5 different sources & Modified Pre-trained AlexNet and a simple CNN & Pre-Trained network achieves a higher accuracy  \\ 
\midrule

\cite{tougaccar2020covid} & 295 \mbox{COVID-19} CXR images and 163 Pneumonia and Normal CXR images & MobileNetV2/SqueezeNet & They use DL models achieve high classification rate\\
\midrule
\cite{ozturk2020automated} & 127 \mbox{COVID-19} CXR images and 1000 Pneumonia and Normal CXR images & DarkCovidNet (CNN) & Provides two approaches, a binary classification and multi-class classification. Binary model performs better and higher\\
\midrule

\cite{rajaraman2020training} & A large dataset containing both \mbox{COVID-19} and Non \mbox{COVID-19} cases from multiple sources  & VGG16, InceptionV3, Xception, DenseNet-121, NasNet-Mobile, etc models compared & VGG16 achieved the highest rate\\ 
\midrule 
\cite{vinod2020data} & 306 \mbox{COVID-19} CXR images and 113 normal CXR images & CNN model using Decision Tree Classifier & The tested method appears to be robust and provides results that are accurate \\ 
\midrule 
\cite{jamil2020automatic} & CXR images of confirmed 150 \mbox{COVID-19} patients from Wuhan \textit{source: Kaggle} & Convolutional Neural Network & The authors are able to get 93\% accuracy \\ 
\midrule
\cite{sethy2020detection} & Multiple datasets with 183 \mbox{COVID-19} CXR images and CXR images of SARS-CoV and MERS & Trained 9 different models for \mbox{COVID-19} & ResNet50 + SVM achieved the highest accuracy\\ 
\midrule  
\cite{toraman2020convolutional} & 231 \mbox{COVID-19} CXR images, and 2100 Pneumonia and normal CXR images \textit{source: Github} & Novel ANN - Convolutional CapsNet & Model provides highly accurate diagnostics with Binary Classification (whether \mbox{COVID-19} or no finding) \\ 
\midrule 
\cite{chowdhury2020can}  & 423 \mbox{COVID-19} CXR images, 3064 CXR images of normal CXR and viral pneumonia & 8 Different CNN Models  & Transfer Learning and Data augmentation Techniques used, CheXNet performs the best \\ 
\midrule
\cite{panwar2020application}  & 192 \mbox{COVID-19} CXR images, and 145 normal CXR images & nCOVnet(VGG-16) & The proposed model is able to achieve high accuracy in predicting \mbox{COVID-19} infected patients from CXR Images \\ 
\midrule
\cite{hurt2020deep} & 10 CXRs from \mbox{COVID-19} confirmed patients in China and USA  & DL U-Net Model & Model shows great promise, with potential use towards early diagnosis for \mbox{COVID-19} pneumonia  \\ 
\midrule 
\cite{ezzat2020gsa} & 126 \mbox{COVID-19} CXR images, 5835 normal and pneumonai CXR images & GSA-DenseNet121-\mbox{COVID-19} (Hybrid CNN using Optimization algorithm) & The proposed model achieves high accuracy in diagnosing, up to 98\% \\ 
\midrule 
\cite{minaee2020deep} & 250 \mbox{COVID-19} CXR images, 4934 Non \mbox{COVID-19} CXR images & ResNet18, ResNet50, SqueezeNet and DenseNet-121 & The models perform well and are tested across multiple parametrs such as Receiver Operating Characteristic (ROC), precision-recall curve, etc. \\ 
\midrule  
\cite{das2020truncated} & Multiple datasets including 162 \mbox{COVID-19} positive CXR images, and Non \mbox{COVID-19} CXR mages   & Truncated Inception Net & The proposed model achieves an accuracy of 99.92\% (AUC 0.99) in classifying \mbox{COVID-19} positive cases   \\ 
\midrule 
\cite{basu2020deep} & 305 \mbox{COVID-19} CXR images, and 822 Non \mbox{COVID-19} CXR images & Transfer Learning method employed on pre-trained models & The authors use Gradient Class Activation Map for detecting where the model focuses more on for classification \\ 
\midrule 

\cite{rahimzadeh2020new} & 180 \mbox{COVID-19} CXR images, and Non \mbox{COVID-19} CXR images & Xception and ResNet50V2 & A concatenated of the two models performs well towards detecting \mbox{COVID-19}  \\ 
\midrule  

\cite{zhang2020covidA} & 318 \mbox{COVID-19} CXR images, and Non \mbox{COVID-19} CXR images & COVID-DA & Propose a Deep Learning model that has a novel classifier separation scheme   \\ 
\midrule 
\cite{apostolopoulos2020extracting} & 455 \mbox{COVID-19} CXR images, and 3450 Non \mbox{COVID-19} CXR images & MobileNetV2 & Training the CNN MobileNetV2 model from scratch proves to get higher accuracy compared to transfer learning techniques  \\
    \bottomrule
  \end{tabular}
   \end{adjustbox}
\end{table*}

\subsubsection{CT-Scan}
\begin{figure*}[h]
\centering
    \includegraphics[scale=.3]{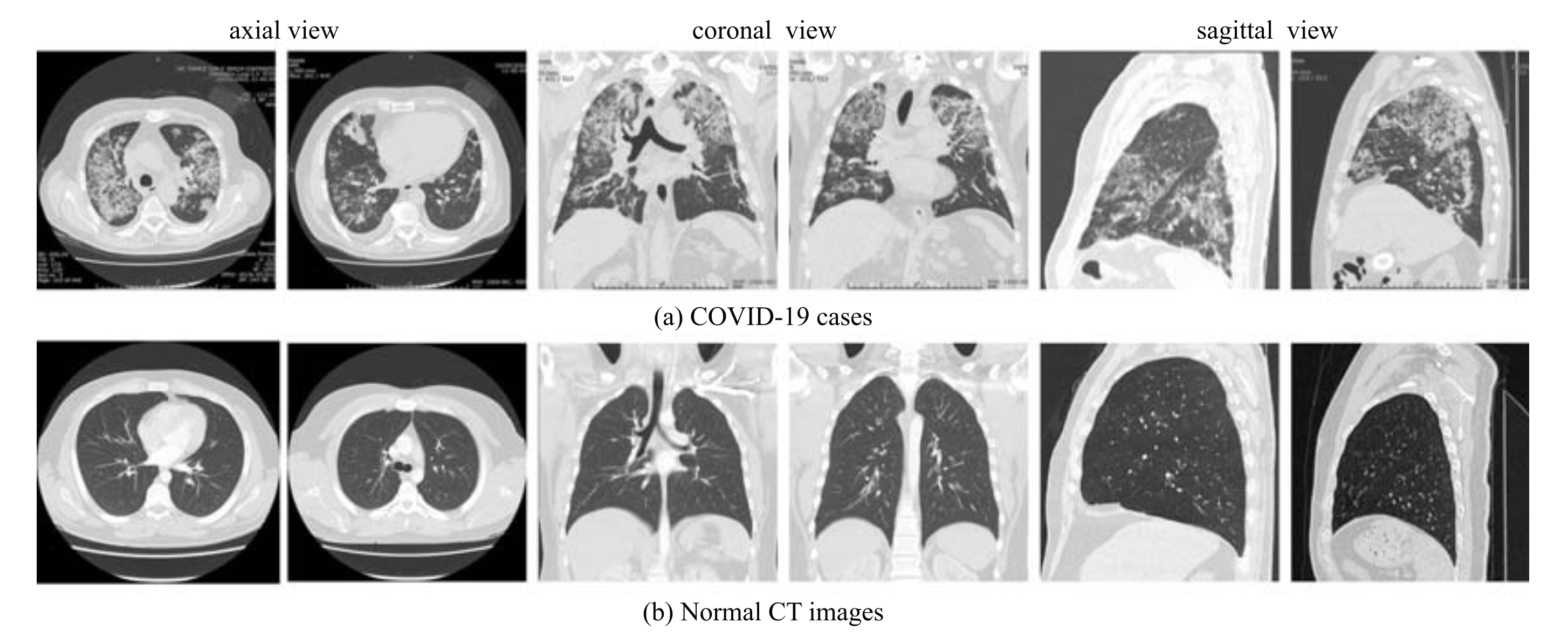}
    \caption{CT-Scan images of \mbox{COVID-19} infected people versus uninfected people \cite{chen2020survey}.}
    \label{image:CTimage}
\end{figure*}

Another applicable medical imaging tool for \mbox{COVID-19} diagnosis is a chest Computed Tomography (CT) Scan which is more accurate in detecting \mbox{COVID-19} cases \cite{choi2020extension}.
Due to respiratory problems of \mbox{COVID-19} which include lung abnormalities, CT-Scan can be specified as the detecting procedure for the early stage of a pandemic while none of the \mbox{COVID-19} symptoms appear in patients \cite{shi2020radiological}. Figure \ref{image:CTimage} demonstrates the CT-Scan images from infected and normal people.

Ardakani et al. \cite{ardakani2020application} implemented a Computer-aided diagnosis system to show the benefits of DL in diagnosing \mbox{COVID-19} using a variety of CNNs which conclude the ResNet-101 as the most precise model. The proposed model can lower the workload of radiology workers as well. Similarly, Li et al. \cite{li2020artificial} developed a DL framework known as \mbox{COVID-19} detection neural network (COVNet) which can differentiate \mbox{COVID-19} from typical types of pneumonia using chest CT-Scan images. 
An ML model for quantitative infection assessment through CT Scan images was modeled by Shan et al. \cite{shan2020lung} suggesting their model is capable of estimating the shape, volume, and the percentage of infection. A Human-In-The-Loop method \cite{holzinger2019interactive} was proposed that involves healthcare workers to intervene with the VB-Net (a modified 3D CNN that combines V-Net and bottle-neck structure) ML model to add more CT-Scan images into the training model to constantly update the model and produce more efficient results. 
Tang, et al. \cite{tang2020severity} introduced another method to detect the severity of \mbox{COVID-19} through assessing quantitative features from CT-Scan images. The use of the Random Forest (RF) model including 500 decision trees along with three-fold cross-validation allows authors to calculate 63 quantitative features of \mbox{COVID-19} like infection volume or lung ratio. However, the model is limited by a binary classification i.e., results can either be severe or non-severe when they should be mild, common, severe, and critical.
Gozes et al. \cite{gozes2020rapid} trained clinical models integrated with ML to detect the virus that achieves high accuracy and is also able to quantify the burden of the disease.  

Similar to the previous section, ML techniques that can be used for detecting \mbox{COVID-19} from CT-Scan images are presented in Table \ref{tab:ctscan}. The list of references includes both published and yet to be peer-reviewed. 

\begin{table*}[ht]
\centering
  \caption{\label{tab:ctscan} ML Research done towards diagnosing \mbox{COVID-19} using CT-Scan datasets.}
 \begin{adjustbox}{width=\textwidth}
 \begin{tabular}{llll}
 \hline
 \textbf{Reference} & \textbf{Dataset} & \textbf{Methods} & \textbf{Remarks} \\ [0.5ex] 
 \hline
 \cite{zhao2020covid}  & Open-sourced COVID-CT \textit{source: Github} & Multi-task and Self-Supervised learning & Clinically useful  \\
 \midrule
 \cite{tang2020severity} & Clinical CT scan images of 176 \mbox{COVID-19} cases & Random Forest / Three-fold cross-validation & The Random Forest model showing promising performance for reflecting the severity of \mbox{COVID-19} \\ 
 \midrule 
 \cite{gozes2020rapid} & Multiple International Datasets of CT scan images (Chinese CDC, Hospitals from China and USA, and Chainz.cn) & ResNet50 & High accuracy in identifying whether \mbox{COVID-19} cases \\ 
 \midrule
  \cite{li2020artificial} & 4356 CT scan images (including \mbox{COVID-19} and Non COVID-19 scans) collected from 6 Hospitals  & Present a Deep Learning model CovNET & The model has the ability to high accuracy in identify \mbox{COVID-19} cases from other lung diseases  \\ \midrule
  
\cite{butt2020deep}  & 618 Clinical CT scan images (including \mbox{COVID-19} and Non COVID-19 scans) & Deep Learning - ResNet18 & Using a location attention mechanism on the model distinguishes \mbox{COVID-19} cases from others with higher accuracy \\ \midrule

    \cite{wang2020deep} & Clinical CT Scan Images from 99 Patients from 3 Hospitals (including \mbox{COVID-19} and Non \mbox{COVID-19} scans)  &  Modified Transfer Learning - Inception Model & Use a fine-tuning technique with pre-trained weights \\ 
    \midrule
    \cite{bai2020predicting} & Clinical CT scan images from 133 Patients from Hospital in China  & Multi Stage - DL Models, LSTM & The model is capable of extracting spatial and temporal information efficiently thereby improving prediction performance \\ 
        \midrule
     \cite{pathak2020deep}  & CT scan images of 413 \mbox{COVID-19} cases and 439 of pneumonia or normal cases  & ResNet50 & The model with transfer learning technique performs better than alternative supervised learning methods \\ 
      \midrule
     \cite{jin2020ai} & CT scan images obtained from 5 Hospitals in China & Various DL learning models (Inception, ResNet50, 3D U-Net++) & Able to deploy the model in 4 weeks overcoming challenges and achieving a high accuracy rate\\
        \midrule
    \cite{shan2020lung}  & 549 CT Scan images obtained from clinics in China & Deep Learning - VB-Net & Introducing a Human-in the loop section to refine automation of each case - the model can segment and quantify infected regions\\  \midrule 
    
    \cite{jin2020development} & Large-scale dataset including 10,250 CT scan images of \mbox{COVID-19} and Non \mbox{COVID-19} scans & UNet - 2D Segmentation DL CNN Model & The DL models outperforms radiologists in diagnostic performance \\  \midrule 
    
    \cite{wang2020noise}  & Clinical CT scan images of 558 \mbox{COVID-19} patients with pneumonia, collected from 10 hospitals  & COPLE-Net + Noise-Robust Dice Loss & The technique presented outperforms standard Noise-Robust loss functions, \\
    
    &  &  &  COPLE-Net and the framework both perform quite highly in segmenting labels for \mbox{COVID-19} pneumonia lesion\\
    \midrule
    
    \cite{song2020end}  & Clinical CT scan images of 83 \mbox{COVID-19} cases and 83 Non \mbox{COVID-19} cases  & BigBiGAN (bi-directional generative adversarial network) & The model achieves high validation accuracy in identifying \mbox{COVID-19} pneumonia from CT images\\ \midrule
    \cite{wu2020jcs}  & Large-scale dataset that include 400 \mbox{COVID-19}, and more Non \mbox{COVID-19} scans  & Classification, Segmentation and Encoder-Decoder Model - Res2Net & Model is highly efficient for Classification and Segmentation  \\ \midrule 
    \cite{zhou2020automatic}  & Multiple datasets that include 473 \mbox{COVID-19} CT scans & UNet & Propose a method to incorporate spatial and channel attention  \\ \midrule 
    \cite{yang2020end}  & Dataset of CT-Scans from  1,684 \mbox{COVID-19} patients & Inception V1 & Validate the model in 3 ways including 10-fold cross-validation achieving high AUC for the validation dataset \\ \midrule
    \cite{yang2020deep}  & Clinical CT scan images including 146 \mbox{COVID-19} cases and 149 Non \mbox{COVID-19} cases & DenseNet & The model classifies \mbox{COVID-19} over CT Images with high AUC \\ \midrule 
    
     \cite{ozkaya2020coronavirus} & 219 CT scan images of \mbox{COVID-19} and 399 CT scan images of normal or other diseases & VGG-16, GoogleNet, ResNet & Support Vector Machine (SVM) was used for binary classification \\ \midrule 
     
      \cite{mobiny2020radiologist}  & 746 CT scan images of \mbox{COVID-19} and Non \mbox{COVID-19} cases; Open-Sourced - \textit{Github} & Capsule Networks (CapsNets), ResNet & Authors present a detail oriented capsule network, implementing data augmentation techniques to overcome lack of data \\ \midrule
      
       \cite{song2020deep} & Clinical CT scans of 88 confirmed patients confirmed with \mbox{COVID-19} from hospitals in China  & DeepPneumonia (ResNet-50) & Model is capable of predicting \mbox{COVID-19} with high accuracy \\ \midrule
       
    \cite{zheng2020deep} & 1,129 Clinical CT scan images for \mbox{COVID-19} detection & UNet, Weakly Supervised DL Network (DeCovNet)  & Accurately predict \mbox{COVID-19} infectious probability without annotating lesions for training  \\ 
\bottomrule
 \end{tabular}
\end{adjustbox}
\end{table*}

\subsection{Chatbots}
\label{chatbots}
Computer programs 
developed to communicate with humans by adopting natural languages 
are called chatbots\cite{shawar2007chatbots}. A chatbot basically can communicate with different users and generate proper responses to those users based on their inputs. 
Recently, the \mbox{COVID-19} pandemic has led to building different chatbots instead of using hotlines as a communication method. This will reduce hospital visits and increase the efficiency of communicating\cite{martin2020artificial}. Generally, chatbots are implemented in order to provide an online conversation with the user by either text or voice displays on web applications, smartphone applications, channels, and so on \cite{cahn2017chatbot}. This conversation can help the user to have a better understanding of his or her situation and gives some hints to users so that he or she can take proper steps.
Chatbots are usually considered as one of the best suited to screen patients remotely without interactions\cite{judson2020case}. The advantages of them include quickly updating information, repetitively encouraging new behaviors such as washing hands, and assisting with psychological support due to the stress caused by isolation and misinformation \cite{miner2020chatbots}. The ML-based chatbots are improved during the training procedure and using more data makes this approach more reliable. 
During the \mbox{COVID-19} pandemic, chatbots are getting more attention in order to provide more details about \mbox{COVID-19} in different stages. A wide variety of chatbots with different languages have been implemented to help patients at the early stage of \mbox{COVID-19}. ``Aapka Chkitsak", an AI-based chatbot developed by \cite{bharti2020medbot} in India, assists patients with remote consultation regarding their health information, and treatments. This application is developed on Google Cloud Platform with the main assistance of Natural Language Processing (NLP) which is compatible with either speech or text. Similarly, Ouerhani et al. \cite{ouerhani2020smart} developed a chatbot (called ``COVID-Chatbot") based on DL model which uses NLP in order to enhance the awareness of people about the ongoing pandemic. COVID-Chatbot is implemented to decrease the impact of the disease during and after the quarantine phase. Another example is Bebot \cite{Bebot2020Chatbot} that provides updated data regarding the pandemic, and also assists patients with symptoms checking. Some other implemented chatbots during this unprecedented time are including Orbita \cite{Orbita2020Chatbot}, Hyro \cite{Hyro2020Chatbot}, Apple's screening tools (website, application, voice command or Siri) \cite{AppleChatbotSiri2020}, CDC's self-checker \cite{CDCchatbot2020}, and Symptoma \cite{Symptoma2020}. A brief description of these chatbots can be found in Table \ref{tab:chatbot}.

\renewcommand{\arraystretch}{1.2}
\begin{table*}[ht]
\centering
  \caption{\label{tab:chatbot} AI chatbots/virtual assistants combating \mbox{COVID-19}.}
 \begin{adjustbox}{width=\textwidth}
 \begin{tabular}{lllll}
 \hline
 \textbf{Reference} & \textbf{AI chatbot/virtual assistant Name} & \textbf{Origin} & \textbf{Company} & \textbf{Function}\\ [0.5ex] 
 \hline
 \cite{bharti2020medbot} & Aapka Chkitsak & India & Academic Research & Remote consultation \\
 
 \midrule
 \cite{ouerhani2020smart} & COVID-Chatbot & Tunisia \& Germany & Academic Research & Enhancing awareness regarding \mbox{COVID-19} \\
 
 \midrule
 \cite{Bebot2020Chatbot} & Bebot &  Japan & Bespoke & Update information \& Symptoms checker  \\

 \midrule
 \cite{Orbita2020Chatbot} & Orbita &  USA & - & Reduce contacts \\
 
 \midrule
 \cite{Hyro2020Chatbot} & Hyro &  Israel & - & Interacting with patients \\

 \midrule
 \cite{Symptoma2020Chatbot} & Symptoma &  Austria & - & Diagnosing by symptoms checking \\

 \midrule
 \cite{Symptoma2020Chatbot} & COVID-BOT & France & Clevy.io & Assist with symptoms by knowledge of government and WHO \\

\bottomrule
 \end{tabular}
\end{adjustbox}
\end{table*}


\subsection{Artificial Intelligence of Things (AIoT)}


In general, applications of the Internet of Things (IoT) \cite{HADDAD2019100129,Sheng-WoT-book,Ngu-IoT} and AI can assist businesses with process automation which would decrease the contacts of humans due to the lower number of people needed \cite{Spiliopoulos2020}. During the \mbox{COVID-19} pandemic, AI and IoT are getting more attention in the healthcare domain where screening and detecting procedures can be done more safely. 
Thermal imaging and social distance monitoring are two main functions that are mainly considered in the screening phase of \mbox{COVID-19}. In fact, the aims of those devices are high-temperature detection, face mask screening, and distance controlling that are discussed in the coming sections.

\subsubsection{Thermal Imaging}
\begin{figure}[h]
\centering
    \includegraphics[scale=.28]{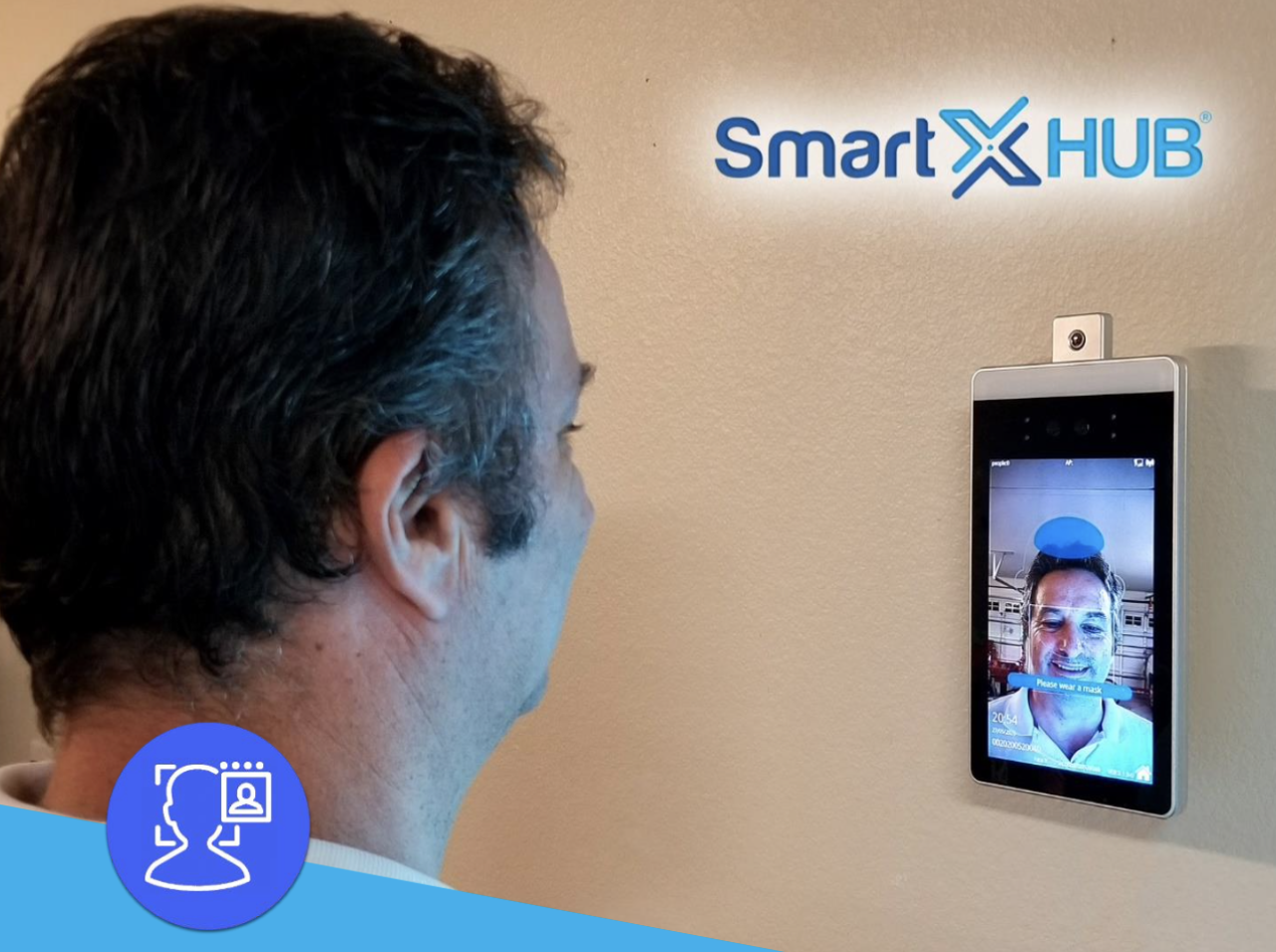}
    \caption{An industrial thermal imaging system enabled with AI \cite{VX2020}.}
    \label{image:SmartX}
\end{figure}

With respect to the IoT thermal screening applications, AI can assist in this area by implementing appropriate algorithms. SmartX\cite{VX2020}, a thermal screening device using infrared thermal imaging and AI face recognition makes screening in crowd buildings or entrances more efficient (see Figure \ref{image:SmartX}). The device captures a visitor's temperature and also checks whether he or she is wearing a face mask. A similar device has been developed in Taiwan for a hospital with the collaboration of Microsoft in order to detect face mask wearings and temperatures. Consequently, any detection of proposed problems can easily be reported to the staff which will maintain an uncontaminated atmosphere\cite{Onag2020}.

\subsubsection{Social Distance Monitoring}
\begin{figure}[h]
\centering
    \includegraphics[scale=.22]{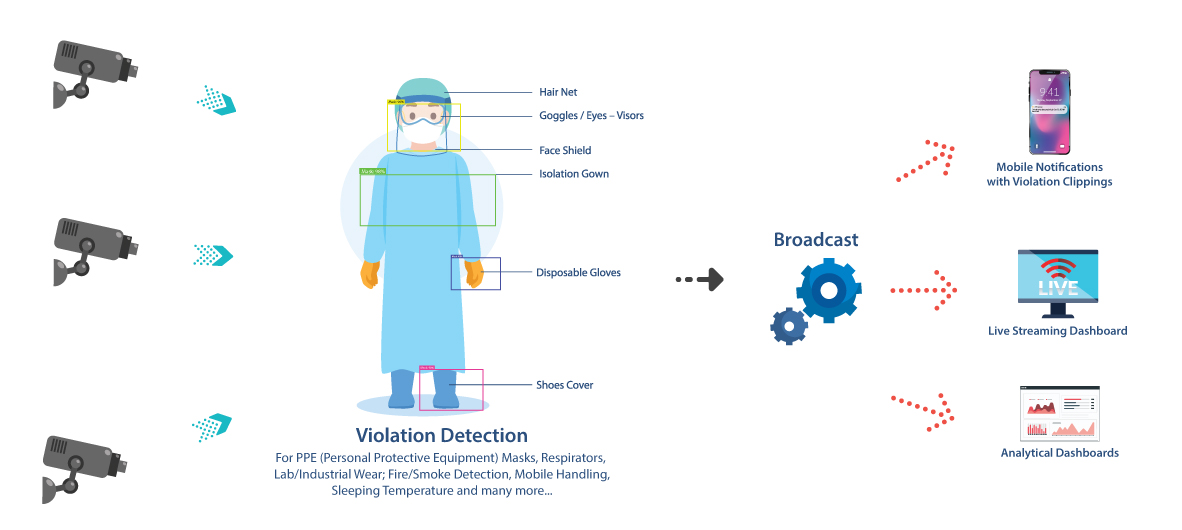}
    \caption{An industrial screening system for monitoring the social distance of people and their personal protective equipment \cite{rayvision2020}.}
    \label{image:RayVision}
\end{figure}

Regarding the necessity of practicing social distancing using IoT devices, AI can implement an automated screening using computer vision methods \cite{Sagar2020}. An instance of such a device is RayVision \cite{rayvision2020} which ensures social distancing and face mask wearing guidelines are followed in the crowd. By using the computer vision techniques, it can monitor people with a live stream on its specific dashboard which allows alerting the authorities in case of any rule-breaking \cite{Waddell2019}. Figure \ref{image:RayVision} represents the process of monitoring using cameras. Similarly, Landing AI\cite{landingAI2020} is another AI-based technology which can detect social distancing violations in real-time. Moreover, a peer-reviewed research \cite{ramadass2020applying} implemented an Unmanned Aerial Vehicle (UAV) or a drone with the ML application in response to the need for maintaining social distance in crowds. Interestingly, masks will be provided by the drone for people who do not wear a mask.

\section{ML Techniques towards predicting and tracking the spread of \mbox{COVID-19}} 
\label{sec:predicting}
As we are currently in the midst of a global pandemic, the ability to predict and forecast the spread of the \mbox{COVID-19} could 
i) help the general population in taking preventative measures, 
ii) allow healthcare workers to anticipate and prepare for the wave of potentially infected patients, 
and iii) allow policymakers to make better decisions regards the safety of the general population. More importantly, the ability to predict the spread of the virus can be used to mitigate and even prevent the spread of \mbox{COVID-19}. 
ML models can be utilized regarding the forecasting of the virus in providing early-signs of the \mbox{COVID-19} and projecting its spread. Also, contact tracing and social media data analysis have shown promising results in  \mbox{COVID-19} spread mitigation \cite{rai2020explainable}. The scope of this section is to provide a review of the research of ML tools and models that can make this possible.

\subsection{Early Signs, Preventing the Spread}
Obtaining early-warning signs for an outbreak of an epidemic could really help towards slowing and mitigating the spread of the virus. It also can encourage societies to take necessary precautionary measures \cite{ganasegeran2020artificial}. In this section, we review the early-warning signs that were made possible using the ML technology.  
The World Health Organization (WHO) made statements about \mbox{COVID-19} being a potential outbreak on the 9th of January 2020. There were AI companies like \textit{BlueDot} and \textit{Metabiota} that were able to predict the outbreak even earlier \cite{naude2020artificial}.
\textit{BlueDot} focuses on spotting and predicting outbreaks of infectious diseases using its proprietary methods and tools. They use ML and NLP techniques to filter and focus the risk of spreading a virus. Using the data from local news reports of first few suspected cases of \mbox{COVID-19}, historical data on animal disease outbreaks, and airline ticket information, they were then able to use their tool to predict a definite outbreak occurring within nearing cities and other regions of China \cite{allam2020artificial}. \textit{BlueDot} had warned its clients about the outbreak on the 31st December 2019, over a week prior to the WHO made any statements about it \cite{naude2020artificial}.
Similarly, \textit{Metabiota} used their ML algorithms and Big Data to predict outbreaks and spreads of diseases, and event severity
\cite{allam2020artificial}. They used their technology and flight data to predict that there will be a \mbox{COVID-19} outbreak in countries like Japan, Thailand, Taiwan, and South Korea.

\subsection{Contact Tracing} 
One of the major approaches for preventing the spread of the virus is tracing the confirmed cases of \mbox{COVID-19} because of the potential spread of the virus through droplets by coughs, sneeze, or talking \cite{CDC2020}. It is recommended that not only the people who have a positive \mbox{COVID-19} test, but also the ones who had been in close contact with the confirmed cases to be quarantined for 14 days. The contact tracing applications are applied all over the world for this purpose with different methods. Basically, it starts after the diagnosis process because the detected case needs to be traced. Most importantly, after the data is collected by those applications, ML and AI techniques will start analyzing for discovering further spread of the disease \cite{Coles2020}. Although the contact tracing applications could be deeply helpful during the pandemic, privacy issues can bring high concerns regarding the surveillance of individuals by some governments as a result of huge amount of the collected data \cite{calvo2020health}. Using the digital footprint data provided by the applications along with ML technology could allow users to identify infected patients and enforce social distancing measures. 

A real-time contact tracing using AI has been applied in South Africa using the Sqreem platform \cite{AIT2020}, which is developed in Singapore in order to track people by the metadata of the device. This information is not including the personal data of the user. If the user enters an infected area, he or she will be contacted by the authorities with respect to the probability of infection. Mostly, contact tracing has been conducted using a major accessible device, which is a smartphone. A variety of smartphone applications enabled with ML or AI have been adopted in order to slow the spread of the virus by tracking and warning regarding the unsafe contacts \cite{kricka2020artificial}. Within the process of development, the very first part would be the consideration of framework, either centralized or decentralized using appropriate technologies such as Global Positioning System, Quick Response codes, and Bluetooth \cite{li2020covid}. ML can enable alerting automatically and analyzing the massive captured data, which would reduce the workforce\cite{Coles2020}. Apple and Google announced that a Bluetooth-based platform for tracking close contacts will be implemented in the upcoming months. This technology will enable higher participation and better communication \cite{panzzarino2020apple}. An application is developed in South Korea in order to capture the areas using location-based information, where a confirmed case went before testing positive for \mbox{COVID-19}. Moreover, text messages will be sent to people who may have been exposed due to the contamination areas\cite{SouthKorea2020App}. 

Regarding the numerous implemented contact tracing applications, Table \ref{tab:ContactTracing} presents some of them that have been implemented in various countries \cite{MITList2020}.

\renewcommand{\arraystretch}{1.2}
\begin{table*}[t]
 \caption{Contact Tracing Applications Combating \mbox{COVID-19}.}
 \label{tab:ContactTracing}
 \centering
 \begin{adjustbox}{width=\textwidth}
\begin{tabular}{lllll}
\hline
  \multicolumn{1}{l}{\textbf{Reference}}&
  \multicolumn{1}{l}{\textbf{Application}}&
  \multicolumn{1}{l}{\textbf{Function}}&
  \multicolumn{1}{l}{\textbf{Origin}}&
  \multicolumn{1}{l}{\textbf{Technology}}
    
  \\\hline

\multirow{2}{*} {\cite{AarogyaSetu2020App}}  &
\multirow{2}{*}
 {AarogyaSetu}
 & Track close contacts of users & \multirow{2}{*} {India} & Bluetooth \\\cline{5-5}\cline{3-3} 
 &     &  Notifying user if captured users are infected &  & GPS location  \\ \cline{5-5}
\hline
\multirow{4}{*}{\cite{Alipay2020App}} & \multirow{4}{*} {Alipay Health Code}  & Track close contacts of users & \multirow{4}{*} {China} & GPS   \\ \cline{5-5}\cline{3-3} 
    & & Display the situation of user by three colors &  & Bank transactions' history      \\\cline{3-3} 
    & & The situations include healthy, in need of short or long quarantine &  &       \\\cline{3-3} 
    & & Tracking traveling information, and body temperature &  &       \\
\hline
\multirow{2}{*} {\cite{BeAware2020App}}  &
\multirow{2}{*}
 {BeAware Bahrain}
 & Track close contacts of users & \multirow{2}{*} {Bahrain} & Bluetooth \\\cline{5-5}\cline{3-3} 
 &     &  Quarantined and self-isolated tracking &  & Location  \\ \cline{5-5}
\hline
\multirow{2}{*} {\cite{COVIDSafe2020App}}  &
\multirow{2}{*}
 {COVIDSafe}
 & Track close contacts of users & \multirow{2}{*} {Australia} & \multirow{2}{*} {Bluetooth} \\\cline{3-3}
 &     &  Notifying user if captured users are infected &  &   \\ 
\hline 
 
\multirow{2}{*} {\cite{CovTracer2020App}}  &
\multirow{2}{*}
 {CovTracer}
 & Track close contacts of users & \multirow{2}{*} {Cyprus} & \multirow{2}{*} {GPS Location} \\\cline{3-3}
 &     &  Notifying user if captured users are infected &  &   \\ 
\hline  

\multirow{2}{*} {\cite{CovidRadar2020App}}  &
\multirow{2}{*} {CovidRadar}
& Track close contacts of users & \multirow{2}{*} {Mexico} & \multirow{2}{*} {Bluetooth} \\\cline{3-3}
 &     &  Notifying user if captured users are infected &  &   \\ 
\hline  
\multirow{2}{*} {\cite{Ehteraz2020App}}  &
\multirow{2}{*}
 {Ehteraz}
 & Track close contacts of users & \multirow{2}{*} {Qatar} & Bluetooth \\\cline{5-5}\cline{3-3} 
 &     &  Notifying user if captured users are infected &  & GPS  \\ \cline{5-5}
\hline
\multirow{2}{*} {\cite{eRouska2020App}}  &
\multirow{2}{*} {eRouska(CZ Smart Quarentine)}
& Track close contacts of users & \multirow{2}{*} {Czech Republic} & \multirow{2}{*} {Bluetooth} \\\cline{3-3}
 &     &  Notifying user if captured users are infected &  &   \\ 
\hline  
\multirow{2}{*} {\cite{GHCovid2020App}}  &
\multirow{2}{*} {GH Covid-19 Tracker App)}
& Track the places an infected user had gone & \multirow{2}{*} {Ghana} & \multirow{2}{*} {GPS} \\\cline{3-3}
&   &  Allow for reporting symptoms &  &   \\ 
\hline  
\multirow{1}{*} {\cite{Hamagen2020App}}  &
\multirow{1}{*} {Hamagen}
&  Track close contacts of users & \multirow{1}{*} {Israel} & \multirow{1}{*} {Location based on API} \\
\hline  
\multirow{2}{*} {\cite{Immuni2020App}}  &
\multirow{2}{*} {Immuni}
& Track close contacts of users & \multirow{2}{*} {Italy} & \multirow{2}{*} {Bluetooth Low Energy} \\\cline{3-3}
 &     &  Notifying user if captured users are infected &  &   \\ 
\hline  
\multirow{2}{*} {\cite{ito2020App}}  &
\multirow{2}{*} {Ito}
& Measure the chance of infection & \multirow{2}{*} {Germany} & \multirow{2}{*} {Bluetooth} \\\cline{3-3}
 &     &  Guide for better safety manner &  &   \\ 
\hline  
\multirow{3}{*} {\cite{maskIR2020App}}  &
\multirow{3}{*} {Mask.ir}
& Track close contacts of users & \multirow{3}{*} {Iran} & \multirow{3}{*} {Bluetooth} \\\cline{3-3}
&     & Provide a map of contaminated areas  &  &   \\ \cline{3-3}
&     & Allow for reporting symptoms  &  &   \\ 
\hline  
\multirow{2}{*} {\cite{myTrace2020App}}  &
\multirow{2}{*} {MyTrace}
& Track close contacts of users & \multirow{2}{*} {Malaysia} & \multirow{2}{*} {Bluetooth Low Energy} \\\cline{3-3}
 &     &  Notifying user if captured users are infected &  &   \\ 
\hline  
\multirow{2}{*} {\cite{StopCovid2020App}}  &
\multirow{2}{*} {StopCovid}
& Track close contacts of users & \multirow{2}{*} {France} & \multirow{2}{*} {Bluetooth} \\\cline{3-3}
 &     &  Notifying user if captured users are infected &  &   \\ 
\hline  
\multirow{3}{*} {\cite{TraceCovid2020App}}  &
\multirow{3}{*} {TraceCovid}
& Track close contacts of users & \multirow{3}{*} {UAE} & \multirow{3}{*} {Bluetooth} \\\cline{3-3}
&     & Accessing to the user's information by government (privacy concern)  &  &   \\ \cline{3-3}
&     & Notifying user if captured users are infected  &  &   \\ 
\hline  
\multirow{3}{*} {\cite{TraceTogether2020App}}  &
\multirow{3}{*} {TraceTogether}
& Track close contacts of users & \multirow{3}{*} {Singapore} & \multirow{3}{*} {Bluetooth} \\\cline{3-3}
&     & Accessing to the user's information by government (privacy concern)  &  &   \\ \cline{3-3}
&     & Notifying user if captured users are infected  &  &   \\ 
\hline  

\end{tabular}
\end{adjustbox}
\end{table*}

\subsection{Forecasting}
Forecasting epidemics centers on tracking and predicting the spread of infectious diseases and viruses. During an epidemic, forecasting methods and models can be trained on epidemiological related data to provide an estimated number of infected cases, patterns of spread that can give healthcare workers guidance on how to prepare appropriately for an outbreak \cite{villela2020discrete}. Previously statistical forecasting tool such as  \textit{Susceptible}, \textit{Infected}, \textit{Recovered} models (SIR Models) have been used to determine the spread of a disease through the population \cite{anastassopoulou2020data}. Recently, with the \mbox{COVID-19} pandemic, using ML approaches for forecasting the spread of \mbox{COVID-19} is getting lots of attention among the research communities.

Hu et al. \cite{hu2020artificial} proposed an unsupervised ML method for forecasting. They used a Modified Auto-Encoder model (MAE) and trained it to predict the transmission of \mbox{COVID-19} cases across 31 provinces or cities in China integrating K-means algorithms to achieve a high prediction accuracy. 
Yang et al. \cite{yang2020modified} used epidemiological data of \mbox{COVID-19} in an SEIR (Susceptible, Exposed, Infectious, Removed) model to predict the spread.
They also presented another interesting approach by pre-training an LSTM (Long Short-Term Memory) model which is a Recurrent Neural Network (RNN) model on data from SARS-CoV to predict the spread of \mbox{COVID-19}. 
Their findings discovered that both models achieved similar results in predicting the number of cases.

Similarly, another ML model with clustering techniques trained on data from the Chinese CDC, internet searches, and news articles create a 2-day ahead real-time forecast about the number of confirmed cases for 32 provinces in China \cite{liu2020machine}. While in some cases there are not enough \mbox{COVID-19} data available for accurate forecasting using ML techniques, Liu et al. \cite{liu2020machine} overcame the lack of data by implementing data augmentation techniques.

Al et al. \cite{al2020optimization} introduced a novel forecasting technique that allows them to predict the number of confirmed cases in China over the next 10 days. The authors combined and modified a Flower Pollination Algorithm (FPA) \cite{abdel2019flower} 
and a Salp Swam Algorithm (SSA) \cite{mirjalili2017salp} to improve and evaluate the optimal parameters for an Adaptive Neuro-Fuzzy Inference System (ANFIS) \cite{karaboga2019adaptive} by creating an FPASSA-ANFIS model that shows greater performance compared to other optimal parameters such as Root Mean Squared Relative Error (RMSE) or Mean Absolute Percentage Error (MAPE).

Alternatively, Rizk et al. \cite{rizk2020covid} integrated algorithms and techniques like Interior Search Algorithm (ISA) and Chaotic Learning (CL) into a Multi-Layer Feed-Forward Neural Network (MFNN) creating a forecasting model called ISACL-MFNN. Combining the two, CL and ISA, approaches improved the overall performance of ISA. The authors retrieved a dataset from WHO that included data from USA, Italy, and Spain between January 2020 and April 2020. They trained their model on this dataset and
the model is then assessed through the similarly aforementioned techniques such as RMSE and MAPE, and more. 

A new ML methodology GROOMs was proposed by Fong et al. \cite{fong2020finding} for forecasting. The authors provided an ensemble of forecasting and polynomial neural techniques that were trained over the limited data to determine the technique that would yield with the lowest prediction error.
%
Ayyoubzadeh et al. \cite{ayyoubzadeh2020predicting} used Liner Regression (LR) and LSTM models to predict the spread of \mbox{COVID-19} in Iran. The authors trained their model on data from Google Trends \cite{GoogleTrends} and Worldometer \cite{WorldPopC19}. They evaluated the model with 10-fold cross-validation and use RMSE as a performance metric.
Ghazaly et al. \cite{ghazaly2020novel} utilized the limited data from WHO about confirmed \mbox{COVID-19} cases and deaths between January 2020 and April 2020 to train a Non-Linear Aggressive Model (NAR) and predict the future cases and deaths that could occur in 9 countries. However, due to not having enough historical data for their model, the authors concluded their network is unable to continue to predict the future.
Over the same time period of those three months \cite{roy2020prediction} Roy et al. implemented their ML techniques to forecast the number of cases for infected, recovered, and deceased cases country-wise and globally. The authors used a type of regression model called the Prophet Prediction Model. Developed by Facebook, this model has a capacity to create a precise time-series forecast that is simple to create and could result in accurate prediction.

Bandyopadhyay et al. \cite{bandyopadhyay2020machine} used a RNN and GRU (Gated Recurrent Unit) model to predict the number of confirmed cases and deaths. The model was trained on data sourced from Kaggle on confirmed cases between January 2020 and March 2020. The results presented in their findings indicate that the RNN is capable of predicting the cases and assessing the severity of \mbox{COVID-19}.
Utilizing the ML techniques, the Global Virus Tracker \cite{Han2020} built up a system embedded the location and symptom risk evaluation for the users. They also developed the chatbot to provide the conversational interface for the same ML-based evaluation. In the proposed system, the county or city level risk, population density and updated virus spreading were incorporated at the moment of the risk evaluation. 



\subsection{Social Media Analysis}
Social media has become a platform where people share pictures, reviews, posts, and exchange stories. A popular social media platform where people may obtain and access news is Twitter. Its users can validate live alerts and obtain information directly through the smartphone application. This is possible as major news outlets, government bodies, community centers, etc.,  all have accounts that they use to share updates on Twitter. Users can also use the platform to share their personal experiences via tweets. It can essentially be considered a form of microblogging for users who just want to share their insight over a certain topic. Tweets can be a form of data that can analyze feedback and obtain public sentiment over certain topics. Over the course of the \mbox{COVID-19} pandemic, people have engaged on twitter to share their experience with regards to testing or lack of testing, social-distancing, and other challenges that people are facing due to the pandemic \cite{cinelli2020covid}. 
In the remainder of the subsection, we review the research dedicated on how ML technology can be used to analyze social media for \mbox{COVID-19} related updates.

The \mbox{COVID-19} was declared a global pandemic in March 2020. 
However, people had already been posting and discussing it over social platforms. Between January 27th and March 26th, 2020 there were 5,621,048 tweets with keywords ``corona virus" and ``coronavirus" according to Karisani et al. \cite{karisani2020mining} when they constructed a dataset for their ML models. 
The dataset consists of both positive and negative tweets shared by users. Using this dataset, the authors were able to implement several ML methods like Logistic Regression (LR), Naive Bayes Classification (NB) to automate the detection of \mbox{COVID-19} positive results shared over twitter by users.
Both LR and NB supervised ML models were also used by Samuel et al. \cite{samuel2020covid} to obtain public sentiment and feedback about \mbox{COVID-19} through users tweets shared over the platform.
As there is an abundance of news and tweets shared over twitter, a sentiment analysis was done by \cite{prabhakar2020informational} addressing the need for filtering it out as there is a potential of misinformation being spread across social platforms. The authors implemented an unsupervised ML topic modelling technique known as Latent Dirichlet Allocation (LDA). 
The use of LDA is done to spot the semantic relationship between words in a tweet and provides a sentiment analysis on whether the tweets are positive and showing signs of comfort or whether they are negative showing discomfort and panic.
In their findings, negative tweets are higher as people are prevalent in anger and sadness towards quarantine and death.

Jahanbin et al.\cite{jahanbin2020using} gathered data from Twitter by searching for \mbox{COVID-19} related hashtags. 
The dataset of tweets is pre-processed and filtered first to reduce the noise by removing irrelevant data. This would allow training a better model for classification. The authors used an evolutionary algorithm Eclass1-MIMO that predicts the morbidity rate in regions.
Mackey et al. \cite{mackey2020machine} introduced an unsupervised ML approach that analyzes tweets by users who may be infected by the virus, recovering from it or the experiences they had related to testing for it. The authors used a Biterm Topic Model (BTM) combined with clustering techniques to determine statistical and geographical characteristics depending on content analysis. 
Obtaining feedback over Twitter and social media outlets will give a live reflection of how the general public is reacting to the pandemic and will allow policymakers in making better decisions.

\renewcommand{\arraystretch}{1.2}
\begin{table*}[t]
 \caption{\label{tab:table-name}Predictive Analysis Tools and Methods Combating \mbox{COVID-19}.}
 \centering
 \begin{adjustbox}{width=\textwidth}
\begin{tabular}{llll}
\hline
  \multicolumn{1}{l}{\textbf{Reference}}&
  \multicolumn{1}{l}{\textbf{Section}}&
  \multicolumn{1}{l}{\textbf{Model and Technology}}&
    \multicolumn{1}{l}{\textbf{Remarks}}
  \\\hline
\cite{allam2020artificial}  &
  \multirow{2}{*}
  {Early Tracking, Prevention}
 & Predictive Analysis tool & Using flight details data and recent outbreaks to predict the spread in nearby countries\\\cline{3-4}\cline{1-1}
\cite{hu2020artificial} &     &  ANN - K-Means Algorithm & Using a MAE to successfully predict 2-day spread  \\\cline{3-4} \hline
\cite{ghazaly2020novel} &
  \multirow{9}{*}  
  {Forecasting}
 & Non-Auto Regressive Neural Network & Prediction Error, due to scarcity of error at the time of Analysis \\\cline{3-4} \cline{1-1}
\cite{liu2020machine} &     &  Augmented ARGONet & Clustering of Chinese Provinces, and getting a 2-day forecast\\\cline{3-4}\cline{1-1}
\cite{fong2020finding} &     &    Polynomial Neural Network (PNN) - GROOMS & Addressing data augmentation and importance of early forecast \\\cline{3-4}\cline{1-1}
\cite{bandyopadhyay2020machine} &     &   RNN & Researching predicting using GRU + LSTM combined models \\\cline{3-4}\cline{1-1}
\cite{siddiqui2020correlation}  &       &  K-Means Clustering Algorithm &   Possible to predict the spread of cases \\\cline{3-4}\cline{1-1}
\cite{al2020optimization}  &       & FPASSA-ANFIS (ANN)  &  Predict a 10-day forecast of the number of cases in China \\\cline{3-4}\cline{1-1}
\cite{rizk2020covid}  &       & ISACL-MFNN & Predict a 10-day forecast of the number of cases in multiple countries  \\\cline{3-4}\cline{1-1}
\cite{ayyoubzadeh2020predicting}  &       & LSTM and LR models & Predict and forecast of the number of cases of \mbox{COVID-19} in Iran \\\cline{3-4}\cline{1-1}
\cite{roy2020prediction}  &       &  Regression Model, Prophet Prediction &   Time-Series Forecasting \\\cline{3-4}\cline{1-1}
\hline
\cite{rao2020identification} &
  \multirow{8}{*}
  {Social Media Analysis}
 & AI Algorithms & Phone based survey to determine whether a person is high-risk, low-risk or contracting the virus \\\cline{3-4}\cline{1-1}
\cite{jahanbin2020using} &     &    Eclass1-MIMO   & Classifying a twitter dataset to determine morbidity in regions \\\cline{3-4}\cline{1-1}
\cite{samuel2020covid} &     &  Natural Processing Language   & Getting public sentiment by classifying tweets \\\cline{3-4}\cline{1-1}
\cite{prabhakar2020informational} &     &  Latent Dirichlet Allocation (LDA) & Algorithms used to spot semantic relationship between words \\\cline{3-4}\cline{1-1}
\cite{zhao2020chinese} &     &  Sentiment Analysis   & Building a visual cluster to highlighting public opinion over pandemic \\\cline{3-4}\cline{1-1}
\cite{mackey2020machine} &     &  Unsupervised ML (biterm topic model) & Attain content analysis by assessing user tweets \\\cline{3-4}\cline{1-1}
\cite{karisani2020mining} &     &  Naive Bayes, Logistic Regression, and more. & Automate detection of positive \mbox{COVID-19} report results through tweets \\\cline{3-4}\cline{1-1}
\cite{schild2020go} &     & Shallow Neural Networks & Training multiple word2vec models to put context to words\\\cline{3-4}\cline{1-1}
\hline
\end{tabular}
\end{adjustbox}
\end{table*}


\section{ML Techniques towards Medical Assistance} \label{sec:medicalAssistance}
As the virus spreads across the world infecting more of the population and with the death toll rising rapidly, efforts are made to develop an effective vaccine or discover a drug for \mbox{COVID-19}. 
The immunologists around the world surged towards studying the symptoms, and the immune responses that infected patients show in relation to combating the virus \cite{gao2017machine}. 
In this section, we review the efforts of the scientific community in using ML techniques regarding understanding the virus \cite{chen2020survey}, how to attack it, and perhaps even be able to find a cure for \mbox{COVID-19}. 


\subsection{Understanding the Virus} 
Analyzing the genomics and proteomics characteristics of a viral disease is an important step to combat the disease. Scientists have been studying the virology of \mbox{COVID-19} which can give the physical and chemical properties, cell entry and receptor interaction, and the overall ecology and the genomics of the virus \cite{lu2020genomic}. 
A genome is the complete genetic information that provides the architecture of a virus and knowing the genome for \mbox{COVID-19} can help in better understanding the transmissibility and infectiousness of the virus \cite{andersen2020proximal}.
The study of proteomics is knowing the proteins of an organism. Identifying the proteins of \mbox{COVID-19} would allow a better understanding of the overall protein structure and discover how the proteins would interact with the drugs \cite{hoffmann2020sars}. Over recent years, there have been remarkable advancements by scientists in interdisciplinary fields of bioinformatics and computational medicine and ML techniques have shown meaningful interpretation towards determining genomics and protein structures of various diseases\cite{lin2017machine}. In this section, we focus on \mbox{COVID-19} and discuss the ML techniques that have been implemented regarding the research of interpreting the genomics and proteomics of that.

\mbox{COVID-19} is an RNA (ribonucleic acid) type of virus from the coronavirus (CoV) family. It is a single-stranded RNA with a large viral genome. These large genomes can have two or three viral proteases. For \mbox{COVID-19} it has 2 proteases, which we will refer to as 3CLpro. \mbox{COVID-19} belongs to the same family as the aforementioned respiratory diseases SARS and MERS \cite{gralinski2020return}. Viruses that belong to this type of family can infect a range of animal species such as camels, cats, cattle, bats, as well as humans \cite{ye2020zoonotic}. They are easily transmittable and can infect a host in one species and transmit it onto another species \cite{ye2020zoonotic}. Multiple findings suggest that the origin of \mbox{COVID-19} in infecting humans is transmitted from bats with an 89\% similarity structure identity to a coronavirus that infects bats (SARS-like-CoVZXC21) \cite{chan2020genomic}. 

Identifying and determining the biological structure at a molecular level of a viral disease is important to develop an effective therapy towards finding a cure for the disease.
However, this process involves a lot of experiments that can take months. Computational ML techniques and models can speed up this process and predict the structure of proteins accurately \cite{DeepMindAlpha}.

The family of coronavirus has multiple classes, and the virus can belong to either the alpha or the beta class of virus. SARS-CoV and MERS both were determined to be beta coronaviruses \cite{CDCVirus}. To determine and classify what specific type of virus \mbox{COVID-19} could be, Randhawa, et al. \cite{randhawa2020machine} used a supervised ML technique with digital signal processing techniques (MLDSP) to evaluate the taxonomy of the virus. MLDSP techniques have previously been used to achieve high accuracy in the classification of other viruses and diseases such as HIV-1 genomes and Influenza \cite{randhawa2019ml}. Using this model, the authors could evaluate that like its predecessor, \mbox{COVID-19} also belongs to the beta coronavirus.
To predict the protein structure of \mbox{COVID-19}, Heo and Feig  \cite{heo2020modeling} used a ML-based method called TrRosetta. This method can be used to predict the inter-residue distance and create structure models for the protein. To do this with higher prediction efficiency, the authors applied molecular dynamics simulation-based refinement. 

Magar et al. \cite{magar2020potential} highlighted the importance of biological structure and protein sequences in combating the virus. The authors developed an ML model to predict how synthetic antibodies inhibit the spread of \mbox{COVID-19}. The model can predict the response of these antibodies by understanding the binding between the antibodies and the viral mutations. 
Training the model on a dataset that includes virus anti-body sequences and the clinical patient neutralization response allowed the authors to predict antibody responses. The authors used ML techniques such as XGBoost, RF, Multilayer Perceptron (MLP), SVM, and LR for their model. 
The XGBoost model achieves the highest accuracy in prediction over an 80\%-20\% split of train and test data.


\subsection{Drug and Vaccine Development}
As \mbox{COVID-19} cases continue to rise numbers of both the infected cases and the death toll, it has become an urgent need to discover a drug that could mitigate these numbers from increasing any further. ML techniques can be used to analyze how drugs react to the viral proteins of \mbox{COVID-19}. We have already seen ML methods and techniques like SVM and Bayesian Classifiers being used for drug discovery and repurposing \cite{chen2018rise}. In this section we review the ML studies and research that had been done about discovering the new drugs or repurposing the currently approved FDA ones. We also review the ML research that has been done regarding the vaccine development. 

\vspace{2mm}
\subsubsection{Drug Development and Repurposing}
An exploratory approach of determining whether commercially available anti-viral drugs can treat or help towards reducing the severity of \mbox{COVID-19} infected patients was presented by Beck et al. \cite{beck2020predicting}. They used a pre-trained ML learning model called Molecule Transformer-Drug Target Interaction (MT-DTI), an interaction prediction model, to predict the binding affinity between \mbox{COVID-19} infected proteins and compounds. The objective of their study was to identify potential FDA approved drugs that may restrain the proteins of \mbox{COVID-19}. MT-DTI is capable of predicting the chemical sequences and amino acid sequences of a target protein without the whole structure information. This is helpful to use as there was limited knowledge on the overall structure of viral proteins of \mbox{COVID-19} initially. Regarding their advantage, the authors used the MT-DTI model to predict binding affinities of 3,410 FDA-approved drugs. 
Similarly, Heiser et al. \cite{heiser2020identification} used proprietary DL techniques for the purpose of drug discovery. They used their model to evaluate how FDA and European Medicines Agency (EMA) approved drugs and compounds would affect human cells, analyzing over 1,660 drugs.

In some cases, various drugs from antiviral to antimalarial could be used to combat \mbox{COVID-19} \cite{moskal2020suggestions}. These drugs used for combating severe illnesses are referred to as ``parents" by Moskal et al. \cite{moskal2020suggestions} in their study. The authors used ML techniques like CNN, LSTM, and MLP analyzing the molecular similarity between these ``parents" drugs and second-generation drugs that could potentially be also used to fight against the virus. The authors introduced the second generation drugs as ``progeny". This study is important in predicting other drugs that may be helpful in this pandemic. It can result in having a larger catalog of drugs, which provides alternative solutions if the ``parent" drug fails to respond.
To convert the molecular structure into a high-dimensional vector space, the Mol2Vec \cite{jaeger2018mol2vec} method was used.

Kadioglu et al. \cite{kadioglu2020identification} identified three viral proteins as targets for their ML approach. They targeted the Spike protein, the nucleocapsid protein, and the 2'-o-ribose-methyltransferase protein. The spike protein acts as a cellular receptor for the host of the virus. The nucleocapsid protein plays a vital role in coronavirus transcription and the overall forming of the genomics of the RNA virus. 
The 2'-o-ribose-methyltransferase protein is an essential protein for coronavirus synthesis and processing. The authors used ML algorithms against the three proteins in predicting how FDA-approved drugs and natural compounds react to the three proteins with such key characteristics.

In \cite{bung2020novo} the authors used a DNN to predict and generate a new small design for molecules that would be capable of inhibiting \mbox{COVID-19} 3CLpro. Targeting 3CLpro can be an essential part with respect to the drugs development for \mbox{COVID-19}.
Alternatively, Zhavoronkov et al. \cite{zhavoronkov2020potential} utilized 28 various types of ML methods such as Autoencoders, Generative Adversial Networks, Genetic Algorithms to predict and generate the molecular structures. Using deep-Q learning networks, Tang et al. \cite{tang2020ai} were able to generate 3CLpro compounds of \mbox{COVID-19}. Being able to successfully predict these protein targets can provide advancements in developing a potential drug for the virus.
Hu et al. \cite{hu2020prediction} created an ML model that predicts the binding between the potential drugs and \mbox{COVID-19} proteins. 


\vspace{2mm}
\subsubsection{Vaccine Development}
Once a virus starts to spread and turn into a global pandemic, there is a very little chance of stopping it without a vaccine \cite{bartsch2020vaccine}. That stands true for \mbox{COVID-19} as well. Historically, vaccination has been the solution to control or slow the spread of a viral infection \cite{weinberg2010vaccine} \cite{harding2018efforts}. It is critical to have a vaccine developed to provide immunity against \mbox{COVID-19} and stop this pandemic. So far, the research for vaccine development of \mbox{COVID-19} is dedicated with three different types of vaccines \cite{chen2020sars}. The Whole Virus Vaccine represents a classical strategy for the development of vaccinations of viral disease. Subunit Vaccine relies on extracting the immune response against the S-spike protein for \mbox{COVID-19} \cite{chen2020sars}. This will refrain it from docking it with the hosts receptor protein \cite{chen2020sars}. The Nucleic Acid Vaccine produces a protective immunological response to fight against the virus by \cite{restifo2000promise}.


At present, there are over 22 vaccines that are at the clinical stage of trials in combating \mbox{COVID-19} \cite{VaccineWHO2020}. The process of developing a vaccine would need to first go through the design stage and then towards the testing and experimental stage on animals and eventually in humans. In this section, we review the implicit research that is done over vaccine development and how ML techniques have been employed.

\begin{table*}[tb!]
  \caption{\label{tab:table-name} Vaccine and Drug Development of \mbox{COVID-19} Using ML Algorithms.}
\begin{adjustbox}{width=\textwidth}
\begin{tabular}{llll}
\hline
  \multicolumn{1}{l}{\textbf{Reference}}&
  \multicolumn{1}{l}{\textbf{Sections}}&
  \multicolumn{1}{l}{\textbf{Model and Technology}}&
\multicolumn{1}{l}{\textbf{Remarks}}
  \\\hline
\cite{ong2020covid} &
  \multirow{7}{*}
  {Vaccine Development}
&  ML Algorithms Logistic Regression, Support Vector Machine, etc & Using a tool called Vaxign to implement Reverse Vaccinology \\\cline{3-4} \cline{1-1}
\cite{liu2020computationally}  &     &   OptiVax, EvalVax, netMHCpan, etc. &  Predicting binding between virus proteins and Host cell Receptors \\\cline{3-4}\cline{1-1}
\cite{ward2020integrated}  &     &  NetMHCPan  &  Create an online tool for visualisation and extraction of \mbox{COVID-19} meta-analysis \\\cline{3-4}\cline{1-1}
\cite{rahman2020epitope}  &     &  Ellipro, multiple ML methods  & Predict the epitope structure \\\cline{3-4}\cline{1-1}
\cite{sarkar2020essential}  &     &  SVM  & Review epitope-based design for a \mbox{COVID-19} vaccine  \\\cline{3-4}\cline{1-1}
\cite{prachar2020covid}  &     &  ANN  & Predicting \mbox{COVID-19} epitopes \\\cline{3-4}\cline{1-1}
\cite{qiao2020personalized}  &     &  DeepNovo, LSTM, RNN  &  Predictive analysis of protein sequences to discover antibodies in patients  \\\cline{3-4}\cline{1-1}
\cite{herst2020effective}  &       &  netMHCpan, netMHC & Use ML techniques to predict peptide sequences \\\cline{3-4}\cline{1-1}
\hline
\cite{heiser2020identification} & 
  \multirow{8}{*}
  {Drug Re-purposing}
 & DL Models - Neural Networks  & Analysing how approved FDA Drugs would work against \mbox{COVID-19} \\\cline{3-4}\cline{1-1}
\cite{beck2020predicting}  &     &   DL - Drug Target Interactions  &   Re-purposing current drugs to find out whether there is an affinity between drug and proteins \\\cline{3-4}\cline{1-1}
\cite{kadioglu2020identification}  &     & Neural Networks and Naive Bayes & Predicting drug interaction between proteins and compounds \\\cline{3-4}\cline{1-1}
\cite{moskal2020suggestions}  &     & CNN, LSTM and MLP models & ML techniques to predict similarities between available drugs to combat \mbox{COVID-19}  \\\cline{3-4}\cline{1-1}
\cite{hu2020prediction}  &     & Fine-Tuning AtomNet based Model   &  Predicting binding between \mbox{COVID-19} proteins and drug compounds \\\cline{3-4}\cline{1-1}
\cite{zhavoronkov2020potential}  &     & Various ML methods & Use RL strategies to generate new 3CLpro structure \\\cline{3-4}\cline{1-1}
\cite{bung2020novo}  &     & Deep Neural Network & Creating small molecule interaction and targeting 3CLpro   \\\cline{3-4}\cline{1-1}
\cite{zhang2020deep}  &   & Deep Learning Models & Provide large scale virtual screening to identify protein interacting pairs \\\cline{3-4}\cline{1-1}
\hline
\end{tabular}
\end{adjustbox}
\end{table*}

Ge et al. \cite{liu2020computationally} used ML techniques to evaluate how small virus strings (called peptides) bind to the human receptor cells. 
The authors created two ML programs OptiVax and EvalVax. Optivax combines different pre-existing programs that evaluate combinations of peptides and receptor cells. A program that OptiVax adopts is NetMHCPan \cite{nielsen2007netmhcpan}, which uses a feed-forward neural network. Thereby, peptide information and receptor cell are input data that generate an affinity score to predict the binding as output. The prediction score is constantly improved as it uses the back-propagation method. 
EvalVax utilizes the genetic ancestry of three different categories of the population, i.e. white, black, and Asian. This data is used as an objective function to discover peptide and receptor pairs for OptiVax.

Similarly, Herst et al. \cite{herst2020effective} in their research about finding a vaccine employed a similar technique that was previously used for combating the Ebola virus. They used ML techniques like Artificial Neural Network (ANN), SVM, netMHC, and netMHCpan to predict potential vaccine candidates.

Ward et al. \cite{ward2020integrated} mapped out the protein sequences of \mbox{COVID-19}. This data was used by the authors for prediction, specificity, and epitope analysis. Epitope is the molecule that adds antibodies attached and is recognized by the immune system. The authors used the ML-based program NetMHCPan to locate the epitope sequences. With this, the authors were able to create an online tool that would aid in epitope prediction, coronavirus homology analysis, variation analysis, and proteome analysis. 
Another study about epitope prediction was presented by Qiao et al. \cite{qiao2020personalized}. They employed DL techniques that are able to predict the best epitope for peptide-based \mbox{COVID-19} vaccinations \cite{qiao2020personalized}. To do this, the authors proposed a DL model that utilizes LSTM and RNN methods to capture sequence patterns of peptides and eventually be able to predict the new mutant antigens for patients.
As an alternative approach to predicting the epitope and protein sequence, an ML based tool called Ellipro was utilized by Rahman et al. \cite{rahman2020epitope}. 
The tool is capable of predicting and presenting a visual view of the protein sequence of the epitope within the structure. Similarly Sarkar et al. \cite{sarkar2020essential} used the SVM method to predict the toxic level of some epitopes.

The study towards epitope prediction continues in the research done by Prachar et al. \cite{prachar2020covid} who employed various techniques such as ANN and Position-Specific Weight Matrices (PSSM) algorithms to predict and verify \mbox{COVID-19} epitopes.
Ong et al. \cite{ong2020covid} introduced a vaccine designing approach referred to as Reverse Vaccinology (RV). The aim of RV is to identify a potential vaccine through bioinformatics analysis of the pathogen genomes. They used Supervised ML classifications such as LR, SVM, K-nearest Neighbor, RF, and Extreme Gradient Boosting (XGB) to train on annotated proteins dataset with the objective of getting high prediction accuracy for the protein candidates, the authors used an ML based tool referred as Vaxign-ML. 

\section{Future Expectations} \label{sec:FutureWorks}
Although we review many of the ML approaches regarding the impact of \mbox{COVID-19} in this paper, there is still an essential need for developing solutions using ML to address the pandemic's complications and challenges. Since there was no adopted method for fighting against \mbox{COVID-19} when it was started, previous ML models regarding infectious diseases (epidemiological models) can be helpful for the early stage of \mbox{COVID-19}. As we discussed, detecting and screening \mbox{COVID-19} using AI and ML techniques can play a key role in combating this pandemic. The combination of technologies like IoT devices with these techniques needs to be expanded for crowd areas including airports, subways or bus stations, and so on. This development would enhance the identification of suspicious cases within lesser both time and contamination. It is important to implement an efficient method of achieving high accuracy of detecting COVID-19 through medical imaging and integrating ML techniques. It is essential to overcome the challenges that are presented by this approach. Challenges such as lack of data and the privacy issue within data collection, false information by media, the limited number of expertise between AI and medical science.
 AI technologies can also assist to implement the following expectations in the future: i) empowering the medical imaging devices using the noncontact automatic image capturing to prevent further infection from the patient to the radiologist or even another patient, and ii) automatically monitoring the patients using intelligent video analysis. As the countries learn more about \mbox{COVID-19}, it is essential to have updated datasets. This can lead to better forecasting by implementing the proper ML models using those datasets. 
In addition to the forecasting concepts, investigating the effects of different social media and their pathways for detecting early sign of possible future pandemic. 


\section{Conclusion} \label{sec:Conclusion}
Machine Learning (ML) models and techniques have vastly been used in plenty of industries over the past decade. Within the healthcare industry, ML has been usually used for screening and diagnosing. In epidemiology area, ML is basically utilized for forecasting and understanding epidemics and diseases. In this paper, we presented a comprehensive survey of how ML applications have been used to fight against \mbox{COVID-19}. We presented the efforts that are taken by the ML research communities to combat this virus across three main phases ``Screening", ``Tracking and Forecasting" and ``Medical Assistance". ML applications for each of these phases are primarily focused as such; ``Screening" intended for diagnosing the virus through medical imaging data (\mbox{COVID-19} related X-Rays and CT-Scans), ``Tracking and Forecasting" towards  forecasting and predicting the numbers of cases and contact tracing, and lastly, ``Medical Assistance" with the aim of understanding the protein sequences and structure of the virus and whether a cure could be found in combating it via a drug or vaccine.
One of the main challenges that researchers face when diagnosing using ML techniques was the lack of relevant data that are made accessible to the public. Lack of data meant researchers had to use techniques like data augmentation, transfer learning, and fine-tuning models to improve prediction accuracy. Though these methods worked well in some cases, more data would make these models more robust. Similarly, forecasting models trained on more data for predicting the spread and number of cases could be more accurate. Regarding developing a vaccine or re-purposing, it is important to have a good understanding of virology, bioinformatics. Additionally, ML is especially important for researchers from different fields to collaborate and integrate their knowledge in order to discover a cure.

\bibliographystyle{IEEEtran}
\bibliography{mybib}

\end{document}